\documentclass{libtex/llncs}
\usepackage{times}
\usepackage{subfigure}

\newcommand{\mitos}{\textsf{Mitos}}
\newcommand{\ROne}{{$PR$}}
\newcommand{\RTwo}{{$OR$}}
\newcommand{\RThree}{{$COR$}}
\newcommand{\RFour}{{$HOR$}}
\newcommand{\ORIF}{{$ORIF$}}

\newcommand{\PGFULL}{{PostgresSQL}}
\newcommand{\PG}{{PSQL}}

\usepackage{epsfig}
\usepackage{libtex/srcltx} 
\long\def\comment#1{}

\newlength{\figtblfootnotemargin}
\newlength{\figtblfootnotewidth}

\newcommand{\SquareOfSize}[2]{
 \fbox{\hsize #1cm \hbox to #1cm{\vbox{#2}}}
}

\newcommand{\putfigureSV}[5]{

    \begin{figure}[htbp]
        \vspace*{-0.1cm}
        \centerline{ \psfig{figure=#1,height=#2}}
        \vspace*{-0.3cm}
        \caption{#4}{\vspace*{0.02cm}  \centerline{ \parbox[t]{10cm}{ #5}}}
        \label{#3}
        \vspace*{-0.4cm}
    \end{figure}
}

\newsavebox{\wholeWidthLine}
\sbox{\wholeWidthLine} {\rule[0.1in]{\textwidth}{.01in}}

\newcommand{\bq}{\begin{quote}}
\newcommand{\eq}{\end{quote}}
\newcommand{\be}{\begin{enumerate}}
\newcommand{\ee}{\end{enumerate}}
\newcommand{\bi}{\begin{itemize}}
\newcommand{\ei}{\end{itemize}}
\newcommand{\bie}{\begin{itemize}\begin{enumerate}}
\newcommand{\eie}{\end{enumerate}\end{itemize}}
\newcommand{\ba}{\begin{array}}
\newcommand{\ea}{\end{array}}
\newcommand{\btbl}{\begin{tabular}}
\newcommand{\etbl}{\end{tabular}}
\newcommand{\bequ}{\begin{displaymath}}
\newcommand{\eequ}{\end{displaymath}}
\newcommand{\bequa}{\begin{eqnarray*}}
\newcommand{\eequa}{\end{eqnarray*}}
\newcommand{\bc}{\begin{center}}
\newcommand{\ec}{\end{center}}
\newcommand{\btab}{\begin{tabbing}}
\newcommand{\etab}{\end{tabbing}}

\def\mpr#1{\ifmmode #1 \else #1 \fi}

\newcommand{\DPATTRNAME}{{\large{{\bf SC}$_{attr}$}}}
\newcommand{\DPISANAME}{{\large{{\bf SC}$_{isA}$}}}
\newcommand{\DPINNAME}{{\large{{\bf SC}$_{in}$}}}

\begin{document}
%
\title{Object-Relational Database Representations \\ for Text Indexing}
\author{Panagiotis Papadakos \ \ \
        Yannis Theoharis \ \ \
        Yannis Marketakis \ \ \\ \  \\
        Nikos Armenatzoglou \ \ \
        Yannis Tzitzikas \\ \ \\
Email: \{papadako, theohari, marketak, armenan, tzitzik\}@ics.forth.gr
}

\institute {Institute of Computer Science, FORTH-ICS, GREECE, and \\
Computer Science Department, University of Crete, GREECE \\
}

\maketitle

\begin{abstract}
    One of the distinctive features of Information Retrieval systems comparing to
    Database Management systems,
    is that they offer better compression for posting lists,
    resulting in better I/O performance
    and thus faster query evaluation.
    In this paper,
    we introduce
    database representations of the index
    that reduce the size (and thus the disk I/Os) of the posting lists.
    This is not achieved by redesigning the DBMS,
    but by exploiting the non 1NF features that existing
    Object-Relational DBM systems (ORDBMS) already offer.
    Specifically, four different database representations
    are described
    and detailed
    experimental results
    for one million pages
    are reported.
    Three of these representations
    are one order of magnitude more space efficient
    and faster (in query evaluation)
    than the plain relational representation.
\end{abstract}

\section{Introduction}

Most information retrieval systems and Web search engines use inverted files,
which have been proven to be very efficient for answering queries \cite{Inv}.
However, the last  years the scope of
services that such systems offer (or should offer)
is  getting wider.
For instance,
they should be able
to handle
structured data
(e.g. Google Base\footnote{http://www.google.com/base}),
structured documents or semi-structured data (e.g. XML),
annotations/tags and multimedia data types.
Furthermore a plethora of new tasks, quite different
from the classical query evaluation task,
are being performed including
data mining algorithms,
machine learning,
facet-based exploration (e.g. \cite{tzitzik08flexplorer,benyitzhak2008bbf}),
collaborative recommendation and filtering.

For these reasons, the index of  an engine
should be easily extensible and able to accommodate various types of data and metadata.
The typical amenities that a DBMS offers
(e.g. declarative query languages, query optimizers),
are very useful
when coping  with multiple types of data (and metadata).
Moreover,
several other techniques and algorithms (e.g. for OLAP)
could be exploited for
enabling services beyond simple search.
In brief,
it is widely accepted, that almost all advanced applications
(including search engines)
need to manage both structured data and text documents \cite{chaudhuri:ida}.
Fortunately, recent work on DB
brings it closer to IR.
For instance
there have been proposed
methods for
ranking query results \cite{agrawal2003ard},
keyword searching in databases \cite{hristidis2003eis,sayyadian2007eks},
computing efficiently top-k queries
\cite{chaudhuri1999etk,ilyas2004stk,RankSQL05},
optimizing text-centric tasks \cite{ipeirotis2006soc},
offering exploration services \cite{chakrabarti2004acq},
and systems that somehow blend
such capabilities
have emerged
(e.g. \cite{conf/cidr/BastW07}).
All these works focus on providing
efficient best-match retrieval services for structured data.
However the management of texts is of prominent importance.
For this reason, in this paper we elaborate on building and managing the index of documents
using a DBMS.
One of the distinctive features of an IR index (e.g. inverted file)
is that it offers better compression
for sparse arrays
resulting in better I/O performance.
In order to alleviate this inefficiency of DBMSs,
in this paper we introduce database representations of the index
that reduce the size (and thus the disk I/Os)
of the representation of the posting lists.
This is not achieved by redesigning the DBMS,
nor by implementing an additional data type,
but by exploiting the non 1NF features that existing ORDBM systems offer.
In brief, Object-Relational DBMSs
extend the relational model to include useful features from object-orientation, e.g. complex types,
and extends relational query languages, e.g. SQL, to deal with these extensions.
\comment{
    ORDBMs use an object oriented database model,
    so that classes, objects and inheritance are directly
    supported by the DB schemas and the query language,
    allowing the extension of the data model with custom data types and methods
}

In this paper, we introduce four different representations (database schemas)
for indexing texts and we report comparative experimental results.
All the experiments have been performed over \mitos\footnote{
http://groogle.csd.uoc.gr:8080/mitos/}.
The index of \mitos\ is based on \PGFULL\ (from now on \PG).
Four different database representations of its index were
tested
for various tasks.
The crux of our findings is that
the support of set-valued attributes by ORDBMSs
can offer significant storage space savings and query evaluation speedup.
To the best of our knowledge this is the first
work that exploits the Object-Relational features of
existing DBMS for the benefit of the index.
There are only few  slightly related works
that are discussed in Section \ref{sec:RW}.
We do not compare
these database representations
with inverted files
because our focus is to identify the more scalable DB representations and not to replace inverted files.
Our findings can significantly speedup
text-centric tasks in settings where a DBMS is already in place.
For instance,
YouTube uses MySQL\footnote{See http://highscalability.com/youtube-architecture},
while there are
Semantic Web repositories, like
SWKM\footnote{http://athena.ics.forth.gr:9090/SWKM/},
that are based on \PG.

The rest of this  paper is organized as follows:
Section \ref{sec:RW} summarizes previous work on DBMS-based IR systems.
Section \ref{sec:Indexer} discusses DBMS indices and 
presents four possible database index representations.
Section \ref{sec:Experimental} reports experimental results.
Finally, Section \ref{sec:Concl} concludes the paper and
identifies issues for further work and research.

\section{Related work}
\label{sec:RW}

One of the first attempts to provide information retrieval functionality such as keyword
and proximity searches by using user defined operators, is described in \cite{Grossman92usingthe}.
Some years later, the first IR system over a DBMS was presented \cite{Grossman95aparallel}.
Relevance ranking queries were implemented using unchanged SQL on an AT\&T DBC-1012 parallel
machine for TREC-3. They found that the DBMS overhead was somewhat high, but tolerable for a large scale machine,
emphasizing that using a DBMS can spread the workload across large numbers of processors.
Recently, several approaches to merge DB's structured data management
and IR unstructured text search facilities have been proposed.
According to \cite{Kim07}, they can be classified in four different categories:

\begin{itemize}
 \item \textbf{Middleware approach}
    This approach integrates DB and IR engines at the application level \cite{chaudhuri:ida}.
    Query evaluation and indexing is provided by the IR engine, while the DBMS manages the documents
    and other metadata.
    According to \cite{Kim07} the basic drawback of this
    approach is the difficulty to synchronize the DBMS document contents and the IR's index.

 \item \textbf{DBMS extension by loose coupling}
    Most DBMS offer extensible architectures using a high level interface, which can be used to integrate
    IR functionalities. Although such extensions can be easily implemented, it is not recommended
    according to
    \cite{Odysseus} when high performance is desired. Systems
    based on this approach, include \emph{PowerDB-IR} \cite{502655}
    (a scalable IR system for frequently changing data sets), \emph{QUIQ} \cite{QUIQ}
    (a collaborative customer support application, where a DBMS holds all the data and an
    external server maintains the index), \emph{TopX} \cite{TopX}
    (an Oracle based engine for XML and plain text data with top-k retrieval)
    and \emph{HySpirit} \cite{fuhr98hyspirit}
    (a hypermedia retrieval engine using probabilistic Datalog).

 \item \textbf{DBMS extension by tight coupling}
    In this approach, new data types and functionality for IR features are integrated into
    the core of the DBMS engine or the reverse
    (IRMS Information Retrieval \& Management System) \cite{Kim07}. Tight coupled systems include
    \emph{Oddyseus} \cite{Odysseus}, an engine build over an ORDBMS engine,
    and \emph{MonetDB/X100} \cite{cornacchia2008fae}, a column oriented storage management based system.

 \item \textbf{DB-IR system from scratch}
    This approach suggests developing new DB-IR architectures from scratch
    \cite{sigmod05,chaudhuri:ida}
    aiming at providing
    structural data independence,
    generalized scoring,
    and flexible and powerful query languages.
\end{itemize}

 The approach that we investigate in this paper falls more into the loose coupling approach.
 No special data types are introduced
 and the retrieval models
 are implemented on top (at a separate API that connects through jdbc to the DBMS).
 However we do exploit the SQL:1999 ARRAY type, allowing the storage of
 a collection of values directly in a column of a table, and the \PG\ (8.2 and above)
 \emph{hstore} data type
 that is useful for storing semi-structural data
 and variable in number fields.
 To the best of our knowledge, the only related work
 is that of \emph{Oddyseus} \cite{Odysseus} and \emph{MonetDB} \cite{cornacchia2008fae}.
 The difference with our work is that
 \emph{Oddyseus} adopts a tight-coupling approach
 where the DBMS is extended with new data types, while
 \emph{MonetDB} implements an inverted file-like data
 structure at the physical layer.
 Specifically Oddyseus adds a B-tree at the posting list of each term
 in order to speedup the lookup of document identifiers and
 the evaluation of multi-word queries.
 However detailed experimental results,
 regarding the space overhead and the speedup of this approach, are not reported.

    In comparison to \cite{papadako08pci},
    this  paper
    (a) contains a detailed discussion of all related works,
    (b) introduces and investigates an additional database representation
        (that yields smaller in size tables),
    (c) reports experimental results over a one order
        of magnitude bigger corpus,
and
    (d) reports experimental results for document-based access tasks.


\section{On DBMS-based Indices}
\label{sec:Indexer}
\subsection{DBMS Limitations}\label{sec:DBMSDrawb}

Roughly, an inverted file comprises entries
of the form
$(t,occ)$
where $t$ is a term
while $occ$
stands for the occurrences of $t$ in the corpus.
Occurrences may comprise only document identifiers,
or also the weight and/or
the positions (exact or block-based) of $t$ in each document.
Term  occurrences occupy most of the space of the index and for this reason
special number encodings \cite{Anh05} are usually employed to reduce the space required.

A straightforward implementation over a relational  DBMS would occupy much more
space than an inverted file.
Consider for example the entry
$(t, \{d_1, d_3, d_5\})$.
That would be represented by
three tuples
$[t,d_1], [t,d_3], [t,d_5]$
resulting in wasted space.
Furthermore,
special number encoding schemes are not currently supported by DBMSs.
Apart from the higher storage space requirements,
we expect the query response time
to be higher for a DBMS based index,
since more I/O's are expected to be needed.
This has been  experimentally verified in \cite{papadakos-2008},
where \mitos\ was found less efficient than Terrier \cite{Ounis-2007}.
However, the adoption of {\em set-valued} attributes
that are offered by  ORDBMSs
can alleviate these problems as we will describe in detail later on.
\comment{
    Specifically  we will study the trade-off,
    between the index size (and query evaluation times)
    and the ability of the index to support multiple access paths,
    e.g. by term  versus by document. In the future we plan to compare
    such DBMS indices with inverted files.
}

\subsection{DBMS Features}\label{sec:DBMSAdv}

Since the scope of services that IR systems and Web search engines
should offer is constantly widening, it is important that they are
based on an {\em easily extensible index}. Using a DBMS index,
the extension of the index schema with additional columns and relations
is rather straightforward.
For instance, the index can be extended with various information,
like users, dates, tags, metadata, in order to
support more sophisticated queries
and retrieval models.
Furthermore, as the physical layer is handled by the DBMS,
the processes of {\em index construction and maintenance} can be simplified (i.e.
there is no need for creating and merging partial indices
in order to construct the index of a big corpus).

Finally, the advances in DBMS for multicore and clustered systems
can  transparently benefit IR systems that are built on top, simplifying
the creation of {\em parallel and distributed systems}.
For instance, \PG\ can take advantage of more than one available system CPUs/cores
(e.g. for dispatching queries),
while the ongoing  project \emph{pgpool-II}\footnote{http://pgpool.projects.postgresql.org/}
works on supporting more advanced distributed query processing features,
such as the dispatching of parts of a query plan to the available CPUs.


\begin{table*}[htb]
\centering

\scriptsize{

\begin{tabular}{|l||c|c|c|}
\hline
   & \multicolumn{3}{|c|}{\textbf{Database Tables}} \\
\hline
\textbf{Repr.}     & \textbf{Document}  & \textbf{Word} & \textbf{Occurrence} \cr
\hline
\hline
\ROne &   [\textbf{\underline{id:{\tt int}}}, \textbf{url:{\tt varchar}}, & [\textbf{\underline{id:{\tt int}}}, 
\textbf{name:{\tt varchar}}, \textbf{df:{\tt int}}]  & [\textbf{\underline{word\_id:{\tt int}}, \textbf{doc\_id:{\tt int}}}, \textbf{tf:{\tt float}}] \\
& \textbf{norm:{\tt float}}, \textbf{rank:{\tt float}}]& & \cr\hline
\RTwo &   [\textbf{\underline{id:{\tt int}}}, \textbf{url:{\tt varchar}}, & \textbf{[\underline{id:{\tt int}}}, 
\textbf{name:{\tt varchar}}, \textbf{df:{\tt int}}]  & [\textbf{\underline{word\_id:{\tt int}}}, \textbf{occur:{\tt Array$\langle$Point$\rangle$}}] \\
& \textbf{norm:{\tt float}}, \textbf{rank:{\tt float}}] & & \cr\hline
\RThree &   [\textbf{\underline{id:{\tt int}}}, \textbf{url:{\tt varchar}}, &  -  & [\textbf{\underline{word\_name:{\tt 
varchar}}},\\
& \textbf{norm:{\tt float}}, \textbf{rank:{\tt float}}] & &   \textbf{occur:{\tt Array$\langle$Point$\rangle$}}, \textbf{df:{\tt int}}] \cr\hline
\RFour &   [\textbf{\underline{id:{\tt int}}}, \textbf{url:{\tt varchar}}, &  -  & [\textbf{\underline{word\_name:{\tt varchar}}},\\
& \textbf{norm:{\tt float}}, \textbf{rank:{\tt float}}] & &   \textbf{occur:{\tt hstore$\langle$text, text$\rangle$}}, \textbf{df:{\tt int}}] \cr\hline
\end{tabular}

}

\small{
\caption{Four Different Database Representations of the Index}\label{tbl:schema}
}
\end{table*}

\subsection{The Indexer of Mitos}
\mitos\ is a recently developed Web search engine in Java, 
that offers  a wide spectrum of functionalities
(for a detailed description see \cite{papadakos-2008}).
Synoptically, \mitos\
is equipped with an advanced stemmer for the Greek language,
offers real time result clustering,
advanced link analysis techniques
and facet-based exploration services \cite{tzitzik08flexplorer}.
\mitos\ adopts the \emph{tf-idf} weighting  scheme and uses \PG\ for managing its index.
For each term it keeps
a)  its document frequency ($df$) in the collection and
b)  its term frequency ($tf$) for each document.
One of the main differences of \mitos\
compared to other search engines,
is that it does not store to the index the positions of term occurrences in documents.
Instead, \mitos\ stores the lexically analyzed extracted text of the crawled pages,
to the filesystem.
When \mitos\ returns the query results to the user,
it parses the stored copies of the texts of the relevant documents,
to find the snippets
with respect to the query terms.
This is needed only for the documents
that lie in the result pages the user will visit.

To compute the answer of a query
the index should provide efficient  {\em term-based} access
(i.e. inverted files).
However there are  other tasks that require {\em document-based} access.
Such tasks include document deletion,
query expansion (retrieve the most highly ranked terms of the top-ranked documents)
and relevance feedback (retrieve the terms of the documents for which the user provided feedback).


\subsection{DB Representations for Occurrences}

Here we introduce four different database representations for the index
(shown in Table \ref{tbl:schema}).
All comprise a relation {\em document},
that  stores for each document
its id, url, norm, and PageRank score.
They only differ on how they store words and occurrences.

\bi
\item[(PR)]{\bf Plain-Relational}

    This is the representation currently in use by \mitos\
    and is like the one used in \cite{Grossman95aparallel,502655,TopX}.
    The relation \emph{word} stores the words,
    their identifiers and their $df$,
    while triples of the form
    $[word\_id, doc\_id, tf]$
    are stored in the  relation \emph{occurrence}.
    The main drawback
    of this representation
    is that  each $word\_id$ is stored
    for each document in which it appears in.
    This redundancy results in high storage space.

\item[(OR)]{\bf Object-Relational}

    This representation exploits the set-valued attributes  supported by
    \PG\ in order to reduce the space occupied by occurrences.
    It exploits the \emph{point} datatype offered by \PG\
    for representing the pairs $\langle doc\_id, tf \rangle$.
    For each  $word\_id$ an array of \emph{points} is stored.
    In this way each $word\_id$ is stored exactly once in the table \emph{occurrence}.

\item[(COR)]{\bf Compact Object-Relational}

    This representation drops the relation \emph{word},
    since $word\_id$ is a
    primary key in both \emph{word} and \emph{occurrence} tables,
    and moves $word\_name$ and $df$
    to {\em occurrence} table.

\item[(HOR)]{\bf HStore Object-Relational}

    This representation is like \RThree, except
    that it uses the \PG\ \emph{hstore} data type instead
    of a \emph{point} array. \emph{hstore} is a data type
    for storing sets of (key,value) \emph{text} pairs in a single
    \PG\ data field. For \RFour\
    the key is the $doc\_id$ and the value is the $tf$.
\ei

\begin{table*}[t]
\centering
\scriptsize{
\begin{tabular}{|l||c|c|c|c|c|c|c|c|}
\hline
\textbf{Repr.}     & \multicolumn{2}{|c|}{\textbf{Document Table}}   & \multicolumn{2}{|c|}{\textbf{Word Table}} & \multicolumn{4}{|c|}{\textbf{Occurrence Table}} \cr
\hline
 & \textbf{Attr.} & \textbf{Type} & \textbf{Attr.} & \textbf{Type} & \textbf{Attr.} & \textbf{Type} & \textbf{Attr.} & \textbf{Type} \cr
\hline
\hline
\ROne   & id    & $B^+$, $Hash$   & name  & $B^+$, $Hash$, $Trie$ & doc\_id   & $B^+$, $Hash$ & $-$ & $-$\cr
\hline
\RTwo   & id    & $B^+$, $Hash$   & name  & $B^+$, $Hash$, $Trie$ & word\_id  & $B^+$, $Hash$ & $-$ & $-$\cr
\hline
\RThree & id    & $B^+$, $Hash$   & $-$   & $-$               & word\_name    & $B^+$, $Hash$, $Trie$ & $-$ & $-$\cr
\hline
\RFour  & id    & $B^+$, $Hash$   & $-$   & $-$               & word\_name    & $B^+$, $Hash$, $Trie$ & occur & with or without $GIN$\cr
\hline
\end{tabular}
}
\small{
\caption{Combinations Between Representations and Indices}\label{tbl:indexes}
}
\end{table*}

\subsection{\PG\ Indices} \label{sec:PSQLIndices}

In order to provide more efficient access paths
to the relations,
we need to build appropriate \PG\ indices.
Regarding \emph{document} table,
the access is done given the $doc\_id$,
i.e., an attribute of integer type.
We have two choices for the index type
we can build on $doc\_id$,
namely either a $B^+Tree$ or $Hash$ index.
Regarding \emph{word} table,
the access is done given the $name$, and we can use a $B^+Tree$ or $Hash$ index.
Furthermore, we could also exploit the \emph{Trie} index,
which has been implemented on top of \PG,
as a part of the SP-GiST index family \cite{SPGiST1,ToTrieOrNotToTrie}.
According to \cite{Eltabakh06}, the \emph{Trie} index offers more than 150\% performance increase for
exact search matches over to \PG\ $B^+Tree$s, and scales better
regarding size. Finally, for the \emph{occurrence} table,
possible choices are either a $B^+Tree$ or $Hash$ index, on
$word\_id$. For the \RThree\ and \RFour\ though, the \emph{word} and
\emph{occurrence} tables have been merged. Since the access is done
given the $name$, we can create either a $B^+Tree$, $Hash$ or $Trie$ index on it.
Moreover in order to accelerate document based access for \RFour, $Generalized Inverted Index (GIN)$
indices can be build on top of the $hstore$ occur attribute.
Unfortunately we could not accelerate document based access for \RTwo\ and \RThree\ ,
since \PG\ does not offer functionality to build indices on top
of arrays.
Table \ref{tbl:indexes} summarizes the
possible combinations.

\comment {
{\bf Gia access by document, isws index sto array attribute na
edine epitaxynsh alla kai extra xwro. Thelei metrhsh)
}
}

\subsection{Bulk Index Creation/Updates}\label{sec:IndexerCreUpd}

It is more than evident,
that the benefits from using a DBMS
are at the expense of the data storage and retrieval efficiency.
Specifically,
the guarantee of the ACID properties,
the concurrency control,
the update of DBMS indices
and their possible reorganization on disc,
may harm the efficiency of the index.
In order to reduce such overheads, we
use the \emph{copy} function of \PG\
during the index creation.
In this manner,
we skip the concurrency control,
as well as several integrity constraints checks,
while at the same time we minimize the I/O's needed
to insert a specific amount of new tuples.
Moreover,
in case we want to add a new document collection
to an existing index,
we first drop the DBMS indices,
then we insert the new tuples,
and finally re-create the indices at the end.
After all documents have been indexed,
for each document $d$
we compute the  norm ($\|\vec{d}\|$) of
its vector ($\vec{d}$)
as defined by the tf-idf weighting scheme,
and store it
in the {\tt norm} field, in order to speed-up query evaluation.

\comment{
Last but not least,
we provide hints to the \PG\ query optimizer
to force it to choose the optimal access paths
(i.e., by taking advantage of the built relation indices)
as well as the optimal query execution plan.}

\subsection{Query Evaluation}\label{sec:IndexerQueries}

Table \ref{tbl:queries}
shows the queries needed according to the
vector space model for each representation,
assuming the query "\textit{information retrieval}" (transformed to
"\textit{informat retriev}" because of
stemming). The query $q_{word}$ is issued to get the $df$ values of the query terms, $q_{occ}$ to get the
$tf$ values of the query terms in the documents they appear in, and $q_{doc}$ to get the norms and ranks
of the corresponding documents.
In \RThree\ and \RFour,
the number of issued queries is decreased by one, since the $df$ values are now
stored in the $occurrence$ table instead of the $word$ table.
These elementary queries can be used
for implementing various ranking methods.
Essentially, they provide the interface that a classical inverted file
exposes to the query evaluation component.

\begin{table}[t]
\centering
{\scriptsize
\begin{tabular}{|l|c|c|c|}
\hline
\textbf{Repr.}   & \multicolumn{3}{|c|}{\textbf{Queries}} \\
\hline
          & \textbf{$q_{word}$} 	& \textbf{$q_{occ}$} & \textbf{$q_{doc}$} \\
\hline
\ROne   & SELECT id, df 			& SELECT word\_id, doc\_id, tf			& SELECT id, norm, rank\\
        & FROM word WHERE 			& FROM occurrence WHERE				& FROM document WHERE\\
        & name IN ('informat', 'retriev')	& word\_id IN (informat\_id, retriev\_id) 	& id IN (doc1, doc2, $...$, docN)\\
\hline
\RTwo   & SELECT id, df 			& SELECT word\_id, occur			& SELECT id, norm, rank\\
        & FROM word WHERE			& FROM occurrence WHERE				& FROM document WHERE \\
        & name IN ('informat', 'retriev')	& word\_id IN (informat\_id, retriev\_id) 	& id IN (doc1, doc2, $...$, docN)\\
\hline
\RThree & 		 			& SELECT word\_name, occur, df			& SELECT id, norm, rank\\
        & -					& FROM occurrence WHERE				& FROM document WHERE \\
        &					& word\_name IN ('informat', 'retriev')		& id IN (doc1, doc2, $...$, docN)\\
\hline
\RThree & 		 			& SELECT word\_name, occur, df			& SELECT id, norm, rank\\
        & -					& FROM occurrence WHERE				& WHERE FROM document\\
        &					& word\_name IN ('informat', 'retriev')		& id IN (doc1, doc2, $...$, docN)\\
\hline

\hline
\end{tabular}
}
\small{
\caption{Queries for each Representation}\label{tbl:queries}
}
\end{table}

\section{Experimental Results}\label{sec:Experimental}

We conducted experiments on a desktop PC
with a Pentium IV 3.4 GHz processor,
2 GB main memory and a single 7200 rpm SATA hard disk,
on top of Linux distribution Ubuntu v8.04, using a 2.6.24 kernel
and the ext3 filesystem (mounted with the default options).
We used \PG\ v8.3.3, configured with 1600 MB as shared\_buffers.
Our collection contained documents of various formats (.html, .pdf, .doc, etc)
including pages crawled  from our
university\footnote{http://www.uoc.gr}
and FORTH\footnote{http://www.forth.gr} domains.
Specifically, it comprises  $1,004,721$ documents,
$216,449$ distinct terms and its total size is approximately 198 GB.
The average size of each document is around 200 KB (due to the large number
of .doc and .pdf files), and the average number of words in each document
is 239.


\subsection{Database Size and Copy Times}\label{sec:DBSizeTimes}

In this section we focus on the storage requirements of the occurrences,
as this is the crucial point
and the main difference between the four representations.
We will use \emph{Object Relational Inverted File} (\ORIF)
to refer to the
\RTwo, \RThree\ and \RFour\ representations, since
they represent occurrences roughly the same
(the only difference is the particular \PG\ data type employed).
For each case,
i.e. \ROne\  and \ORIF,
we consider two different settings, depending on whether the
positions of the occurrences are stored or not in the index.
By adopting the notations described in  Table \ref{tbl:SizeNotations},
the size of each representation can be estimated as follows:

\begin{itemize}
 \item \ROne\ (without positions):
    $N_d  (3 f + t)$\\
    $N_d$ is multiplied
        by $(3 f + t)$, because  for each occurrence
        we have to keep a tuple
        containing the corresponding \emph{word\_id}, \emph{doc\_id} and \emph{tf}.
        Recall that $t$ is the tuple overhead of the DBMS
        and is independent of the attributes of the tuple.

 \item \ROne\ (with positions):
    $N_d  (3 f + t) + N  (3  f + t)$\\
    $N_d$ is multiplied by $(3  f + t)$
    for the same reason as in the \ROne\ (without positions) case.
    In addition, $N$ is multiplied by $(3  f + t)$,
    since for each occurrence we have to keep
    a tuple
        containing the corresponding
        \emph{word\_id}, \emph{doc\_id} and \emph{position}.

\item \ORIF\ (without positions):
    $W  (f + t) + N_d 2 f $\\
    We have to store $W$ tuples holding for each word the \emph{word\_id} and
    $N_d$ pairs of \emph{doc\_id} and \emph{tf}.
    No extra $t$ has to be payed as they are stored in the same tuple.
\
 \item \ORIF\ (with positions):
    $W (f + t) + N_d  2  f + N  f$\\
    Again we have to store $W$ tuples holding for each word the \emph{word\_id},
    and
    $N_d$ pairs of \emph{doc\_id} and \emph{tf} in the same tuple.
    Moreover, we have to store in the same tuple $N$
    fields holding the positions where terms appear in.

\end{itemize}

\begin{table}[t]
\centering
{\scriptsize
\begin{tabular}{|l|p{55mm}|}
\hline
{\textbf{Notation}} & {\textbf{Definition}}\cr
\hline
\hline
    $N$ & number of word occurrences in the entire collection\cr
\hline
    $D$ & number of documents\cr
\hline
    $N_d$ & $\sum_{i=1..D}{w(d_i)}$ where $w(d_i)$ is the number of distinct words in document $d_i$.\cr
\hline
    $W$ & number of distinct words of the entire collection\cr
\hline
    $t$ & tuple size (the space overhead for a tuple in a DBMS)\cr
\hline
    $f$ & field size of a tuple\cr
\hline
\end{tabular}
}
\small{
\caption{Size Notations}\label{tbl:SizeNotations}
}
\end{table}

Hereafter we focus on the case where we do not store positions.
It is clear
that \ORIF\ occupies less space than \ROne.
The inequality \ORIF $<$ \ROne\  yields:
$W (f + t) + N_d  2  f   < N_d  (3  f + t) \Leftrightarrow$ 
$W (f + t)               < N_d f + N_d  t \Leftrightarrow$
$W (f + t) < N_d (f + t) \Leftrightarrow$
$W                       < N_d$
which is always true since every word in
$W$ (where $W$ denotes the vocabulary, not its cardinality)
will appear in at least one document
($W = N_d$ if each distinct word appears in exactly one document and each document
contains exactly one distinct word).
Regarding the lower and upper bounds of $N_d$
(recall that $N_d = \sum_{i=1..D}{w(d_i)}$),
as $1 \leq w(d_i) \leq W$,
 it follows that $D \leq N_d \leq D W$.
Moreover if we assume that each document has $w_{avg}$
distinct words,
then
$N_d = w_{avg} D$.
In our collection $w_{avg}$ is 239 words,
so \ROne\ is expected to occupy much more space than \ORIF.
In  the case  that we also store term positions,
it is clear  that \ORIF\ will again require less space,
as in \ROne\ we use $N  (3 f + t)$ to store the positions instead of
just $N * f$ in \ORIF.

Regarding the physical database size for each representation,
we consider
that the \PG\ storage requirement for string types
is 4 bytes plus the actual string size,
while the storage requirement for integers and floats
(considering the {\tt int4} and {\tt float4} types respectively)
is 4 bytes and the size of type {\tt point} is 16 bytes\footnote{\PG\ version 8.3 supports
arrays of composite types. Thus we could create a composite type
(holding an {\tt int4} and a {\tt float4} (8 bytes) instead of the {\tt point} type),
reducing the memory size of the array to half}. In addition,
the storage cost per tuple is 40 bytes,
due to an internal id generated to
identify the physical location of a tuple within its table,
i.e. $t=40$ bytes.

\begin{table*}[t]
\centering
\scriptsize{
\begin{tabular}{|l||c|c|c|c||c|c|c|c||c|}
\hline
\textbf{Repr.} & \multicolumn{4}{|c||}{\textbf{Number of Pages (8KB each) per Table}} & \multicolumn{4}{|c||}{\textbf{Number of Tuples per Table}} &  \textbf{Time}\cr
\hline
 &    \textbf{Document}  & \textbf{Word} & \textbf{Occurrence}   & \textbf{Total} & \textbf{Document}  & \textbf{Word} & \textbf{Occurrence}   & \textbf{Total} & \textbf{Copy}\cr
\hline
\hline
    \ROne   & 35,654  & 1,278  &  1,301,657   & 1,338,589($\sim$10.7GB)  & 1,004,721 & 216,449 & 240,806,511 & 242,027,681 & $\sim$2.8 days\cr
\hline
    \RTwo   & 35,654  & 1,278  & 28,577     & 65,509($\sim$524MB)    & 1,004,721 & 216,449 & 216,449 & 1,437,619 & $\sim$24 min.\cr
\hline
    \RThree & 35,654  & $-$   & 28,860     & 64,514($\sim$516MB)    & 1,004,721 & $-$ & 216,449 & 1,221,170 & $\sim$25 min.\cr
\hline
    \RFour  & 35,654  & $-$   & 24,106     & 59,760($\sim$478MB)    & 1,004,721 & $-$ & 216,449 & 1,221,170 & $\sim$35 min.\cr
\hline
\end{tabular}
}
\small{
\caption{DB Tables Size in Pages (8 Kb) and Indexing Times}\label{tbl:DBSizeTimes}
}
\end{table*}

The sizes of the tables for each representation
that correspond to our collection
are given in
Table \ref{tbl:DBSizeTimes}.
Notice that the sizes of \ORIF\
are significantly smaller (more than one order of magnitude).
Specifically \ROne\ occupies 10.7 GB, while \ORIF\
occupy around 0.5 GB.
This means that the storage space of \ORIF\
is roughly  0.25\% of the total collection size.
As a consequence the times to
copy the tables are significantly smaller for \ORIF, in comparison
to \ROne, offering a much more scalable solution, as far as indexing time and size are concerned.
Specifically, \ROne\
needs almost 3 days to copy the tables, while the other representations need only 30 minutes
or less.

\begin{table*}[tbh]
\centering
\scriptsize{
\begin{tabular}{|l||c|c|c|c||c|}
\hline
\textbf{Repr.}     & \multicolumn{4}{|c|}{\textbf{Database Table Indices}}   & \multicolumn{1}{|c|}{\textbf{Time}}  \cr
\hline
    &    \textbf{Document}  & \textbf{Word} & \textbf{Occurrence}  & \textbf{Total} & \textbf{Index Creation} \cr
\hline
\hline
\ROne\ using $Hash$         & 4,666  & 2,050  &  1,466,665   &  1,473,381 ($\sim$11.7 GB)  & 77,793.7s ($\sim$21.5 hours)\cr
\hline
\ROne\ using $B^+Tree$      & 2,208 & 720 & 528,421 & 531,349 ($\sim$4.2 GB)    & 2,150s ($\sim$35 min)\cr
\hline\hline
\RTwo\ using $Hash$         & 4,666   & 2,050  & 1,168      & 7,884 ($\sim$63 MB)      & 13.0s\cr
\hline
\RTwo\ using $B^+Tree$      & 2,208   & 720  & 478      & 3,406 ($\sim$27 MB)      & 11.6s \cr
\hline\hline
\RThree\ using $Hash$       & 4,666   & $-$   & 2,050      & 6,716 ($\sim$53 MB)      & 11.6s\cr
\hline
\RThree\ using $B^+Tree$    & 2,208   & $-$   & 720      & 2,928 ($\sim$23 MB)      & 6.4s \cr
\hline\hline
\RFour\ using $Hash$        & 4,666   & $-$   & 2,050      &  6,716 ($\sim$53 MB)     & 6.9s\cr
\hline
\RFour\ using $B^+Tree$     & 2,208  & $-$   & 720 & 2,928 ($\sim$23 MB)       & 6.7\cr
\hline
\end{tabular}
}
\small{
\caption{Indices Size in Pages (8 KB) and Creation Times}\label{tbl:PSQLIndexesSizeTimes}
}
\end{table*}

\subsection{Indices Size and Creation Times}\label{sec:PSQLIndexesSizeTimes}

The sizes of the \PG\ indices for each representation are shown in Table \ref{tbl:PSQLIndexesSizeTimes}.
Again the space difference between the representations is more than one order of magnitude.
This is also reflected
to the \PG\ index creation times.
Specifically the process takes some seconds in \ORIF,
and roughly a day for \ROne.
We could not evaluate the \emph{Trie} index,
as it only accepts words with latin characters and
our test collection  mainly contained greek documents.
In addition we could not evaluate \emph{GIN} indices, in order to
accelerate document based access\footnote{GIN index creation query was running
for 3 days, before we canceled it due to time limitations}.
In general, the results show that $B^+Tree$ indices occupy half of the size of  $Hash$ indices.

\subsection{Query Evaluation Times}\label{sec:QETimes}

To measure query evaluation times, we adopted the following scenario:
for each of the four representations and for each \PG\ index combination, we:
a) execute all the queries of the corresponding representation
with 1, 2, 3 and 4 terms, b) repeat the above queries 10 times and
c) calculate average times.
We do not include the time to receive/scan the results.
The terms contained in the above queries were different
(for each query and for each iteration of the experiment)
and they were selected based on their $df$.
Specifically, we selected frequently occuring terms
with a $df$ value about 300,000.
The big $df$ number implies big overhead to the DBMS.
The crash we had encountered in our previous experiments \cite{papadako08pci} using \PG\ 8.0,
due to the large number of $doc\_id$s passed in the IN list of the $q_{doc}$ queries,
was solved after upgrading to \PG\ 8.3.
The aforementioned times were gathered through
the {\tt Aggregator}\footnote{http://www.csd.uoc.gr/$\sim$andreou}
toolkit which is written in Java.
This means that the measured times include the overhead of the JDBC driver (version 8.3-603 JDBC 4),
an overhead that also exists in the \mitos\ engine.

\begin{table*}[t]
\centering
\scriptsize{
\begin{tabular}{|p{6mm}||c|c|c|c||c|c|c|c||c|c|c|c||c|c|c|c|}
\hline
\textbf{Repr.}     & \multicolumn{4}{|c||}{\textbf{1 term}}   & \multicolumn{4}{|c||}{\textbf{2 terms}} & \multicolumn{4}{|c||}{\textbf{3 terms}} & \multicolumn{4}{|c|}{\textbf{4 terms}} \cr
\hline
    & \textbf{$q_{w}$} & \textbf{$q_{occ}$} & \textbf{$q_{doc}$}  & \textbf{$tot$} & \textbf{$q_{w}$} & \textbf{$q_{occ}$} & \textbf{$q_{doc}$} & \textbf{$tot$} & \textbf{$q_{w}$} & \textbf{$q_{occ}$} & \textbf{$q_{doc}$} & \textbf{$tot$} & \textbf{$q_{w}$} & \textbf{$q_{occ}$} & \textbf{$q_{doc}$} & \textbf{$tot$}\cr
\hline
\hline
\ROne\ $(H)$  & 64 & 54,418 & 3,949 & \textbf{58,431} & 90 & 94,544  & 8,689 & \textbf{103,323} & 84 & 148,220 & 9,620 &\textbf{157,924}& 95 & 202,253 & 12,778 & \textbf{215,126} \cr
\hline
\ROne\ $(B^+)$ & 88 & 33,633 & 5,299 & \textbf{39,020} & 74 & 63,330 & 7,750 &  \textbf{71,154} & 88 &  99,808 & 9,456 & \textbf{109,352}& 101 & 131,317 & 12,073 &  \textbf{143,491} \cr
\hline\hline
\RTwo\ $(H)$ & 16 & 846 & 1,952 & \textbf{2,814} & 18 & 1,650 & 4,744 & \textbf{6,412} & 24 & 2,443 & 7,402 & \textbf{9,869} & 32 & 3,153 & 10,128 & \textbf{13,313} \cr
\hline
\RTwo\ $(B^+)$ & 6 & 766 & 2,143 &\textbf{2,915} & 7 & 1,490 & 5,616 &\textbf{7,130} & 7 & 2,337 & 8,084 &\textbf{10,428} & 5 & 3,018 & 10,053 & \textbf{13,076}\cr
\hline\hline
\RThree\ $(H)$ &$-$& 856 & 3,391 & \textbf{4,247}&$-$& 1,618 & 5,777 & \textbf{7,395} &$-$& 2,419 & 7,902 & \textbf{10,321} &$-$& 3,292 & 9,873 & \textbf{13,165}\cr
\hline
\RThree\ $(B^+)$ & $-$& 798& 4,447& \textbf{5,245}&$-$& 1,529& 5,605& \textbf{7,134}&$-$& 2,349& 7,982& \textbf{10,331}&$-$& 3,085& 10,237& \textbf{13,322}\cr
\hline\hline
\RFour\ $(H)$ & $-$ & 1,023& 3,944& \textbf{4,967}&$-$& 1,280 & 5,637&\textbf{6,917} &$-$& 1,949& 8,023& \textbf{9,972} &$-$&2,424 & 9,9993&\textbf{12,417}\cr
\hline
\RFour\ $(B^+)$ & $-$ &127 & 3,208& \textbf{3,335} &$-$&255 &5,627 &\textbf{5,882} &$-$& 319& 8,007& \textbf{8,326}&$-$&518 &10,098 &\textbf{10,616}\cr
\hline
\end{tabular}
}
\small{
\caption{Query Evaluation Times (msec)}\label{tbl:QETimes}
}
\end{table*}

As one can observe from the times reported in Table \ref{tbl:QETimes},
\ORIF\ representations are one order of
magnitude more efficient than \ROne, due to the efficiency in occurrence table.
More precisely, \ORIF\ are approximately
20 times faster than \ROne\ for all queries, although
the \ORIF\ index
is only an order of magnitude (see Table \ref{tbl:DBSizeTimes})
smaller than \ROne\ index.
This is due to the fact that
\ORIF\ indices, fit in main
memory, so every page that is fetched in memory, is constantly kept there.
Comparing \ORIF\ representations, we observe that \RTwo\ and \RThree\
have an identical performance, while \RFour\ is slightly faster, especially
when using a $B^+Tree$ index.
A common behavior for all representations is the
slow $q_{doc}$ query times, which is actually the bottleneck for \ORIF.
This is due to the long IN list of $doc_{id}$s.
We found that passing more than 250,000 $doc\_id$s makes the $q_{doc}$
query too slow. We tackled this problem by dividing the IN list in blocks of 250,000 $doc\_id$s and
submitting one query for each block of the list.
Subsequently we summed  the gathered times.
In the future we plan to investigate whether
we can reduce the overhead of such queries by using temporary tables.
Regarding the DBMS indices,
we can conclude that $B^+Tree$ indices are the best choice
since they provide equivalent or slightly better
performance to $Hash$ indices,
while occupying half the storage space.

\subsection{Query Expansion Times}\label{sec:QExTimes}

To measure query expansion times
we used the  following scenario:
for all terms in the top-5 results of a given query,
we compute the sum of their $tf$s in these documents
and suggest to the user the 5 terms with the highest sum.
Unfortunately, the gathered times were unacceptable for \ROne\ (almost 16 hours),
since no index over $doc\_id$ existed
and \PG\ performed a slow sequential scan over a table of
240,000,000 tuples.
This task is much faster for \ORIF\ (19.8 minutes), but again expensive,
since an index over the \emph{array} and \emph{hstore} values, was not build.
An approach to speed-up such tasks, is to store also a {\em direct index}
(keeping for each $doc\_id$ the set of $word\_id$s it contains).
This index can be represented
in an \ORIF-like representation.
We expect
the total size
of the two \ORIF\ representations (inverted and direct)
to be less than that of \ROne, and at the same time should provide faster
term-based and document-based acess services.

\putfigureSV{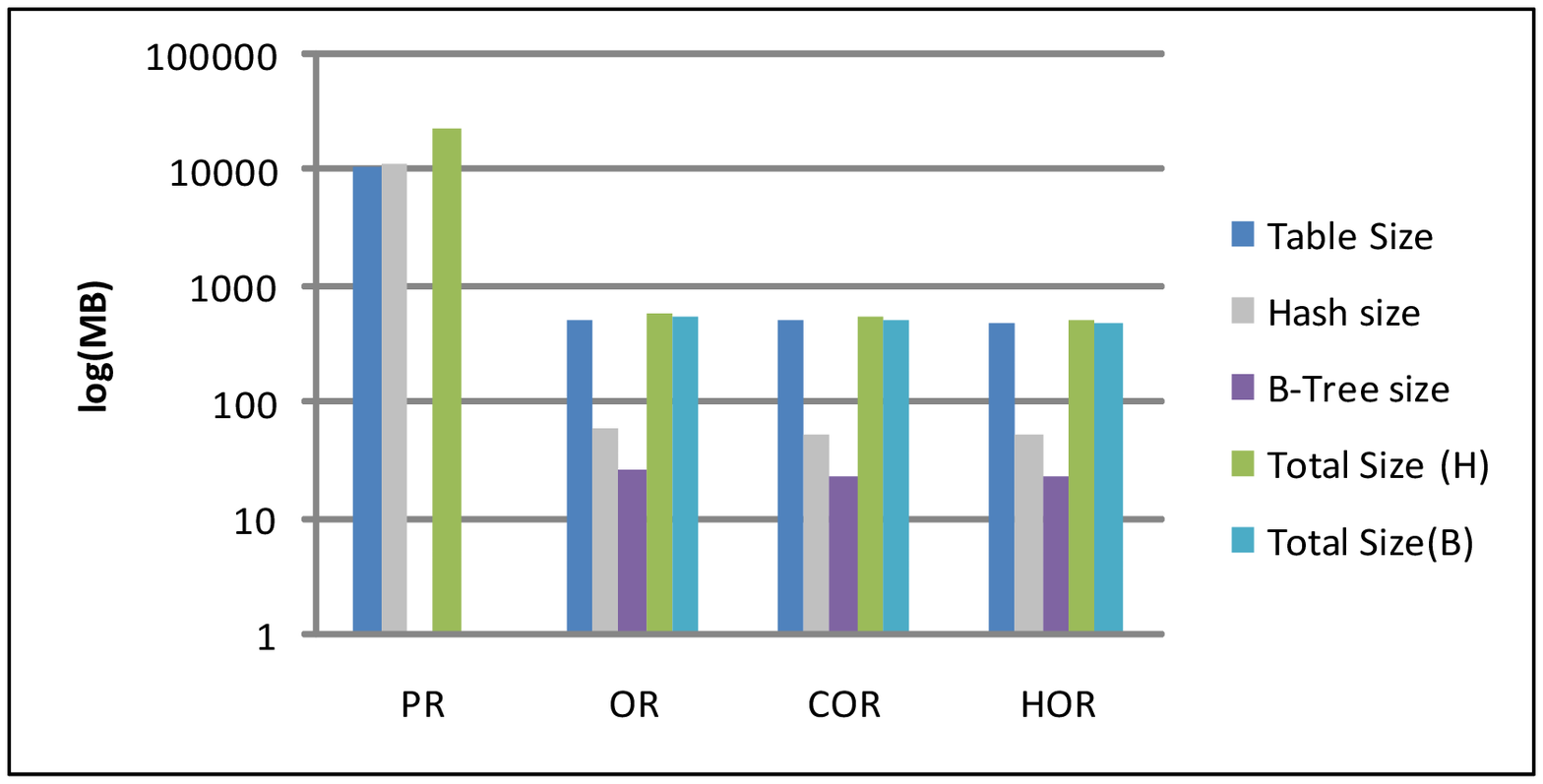}{43mm}{fig:sizes}
  {Size of Tables and Indices (in log scale)}{}

\section{Conclusion}
\label{sec:Concl}

\begin{figure}[htp]
     \centering
     \subfigure[Times to Copy DB Tables (in log scale)]{
          \label{fig:indexCopyTime}
          \includegraphics[width=54mm]{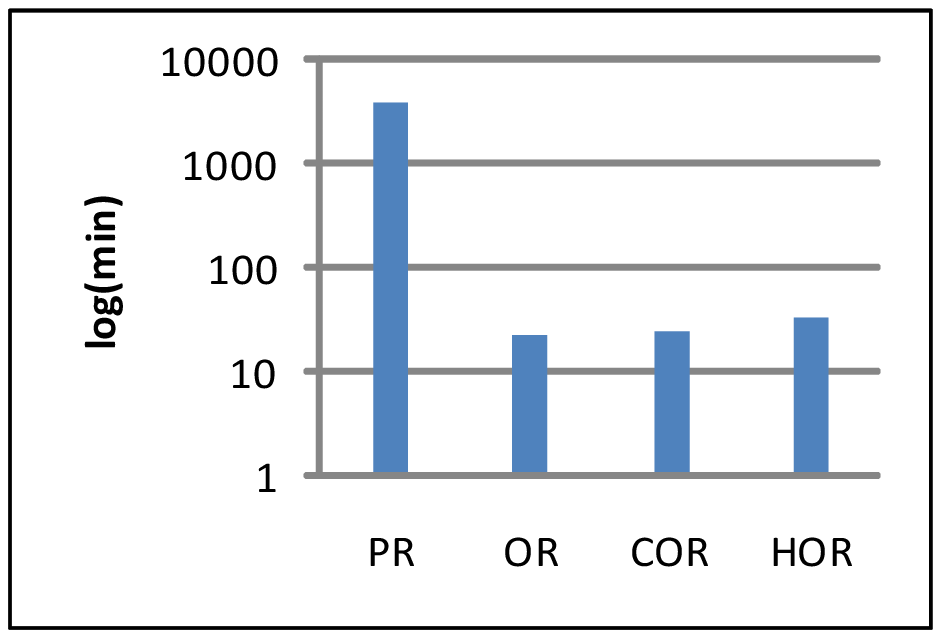}}
     \hspace{.1in}
     \subfigure[Creation Times of \PG\ Indices (in log scale)]{
          \label{fig:dbCreationTime}
          \includegraphics[width=56mm]{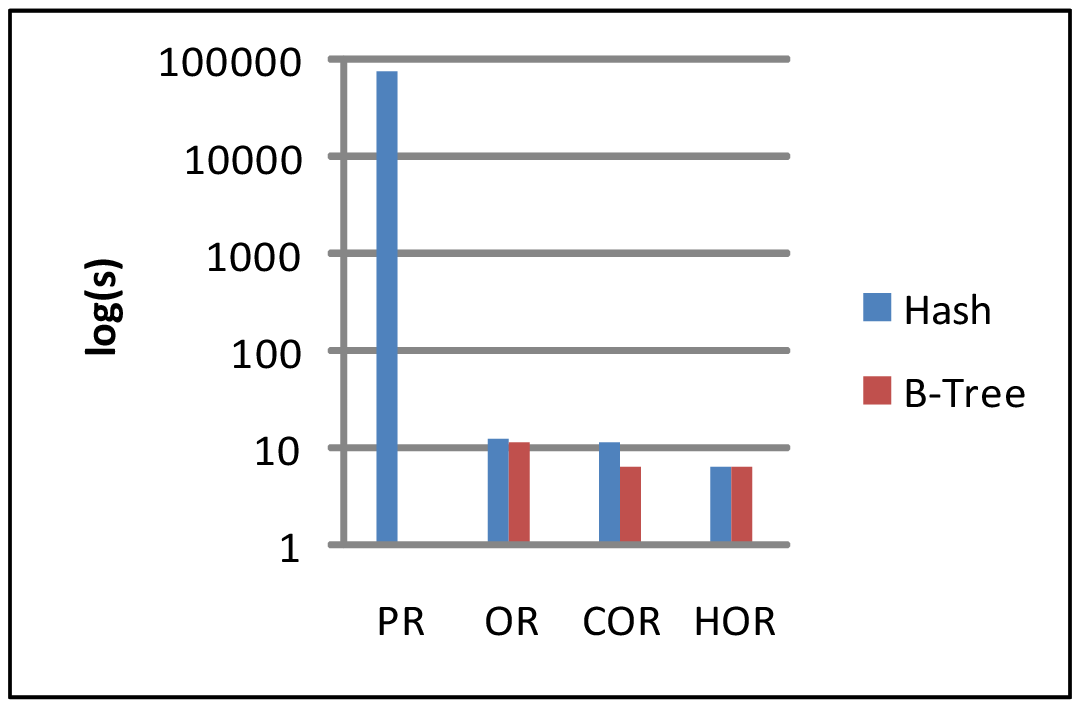}}
     \hspace{.1in}
     \subfigure[Query Evaluation using $B^+Trees$]{
          \label{fig:queryEvaluationBTree}
          \includegraphics[width=56mm]{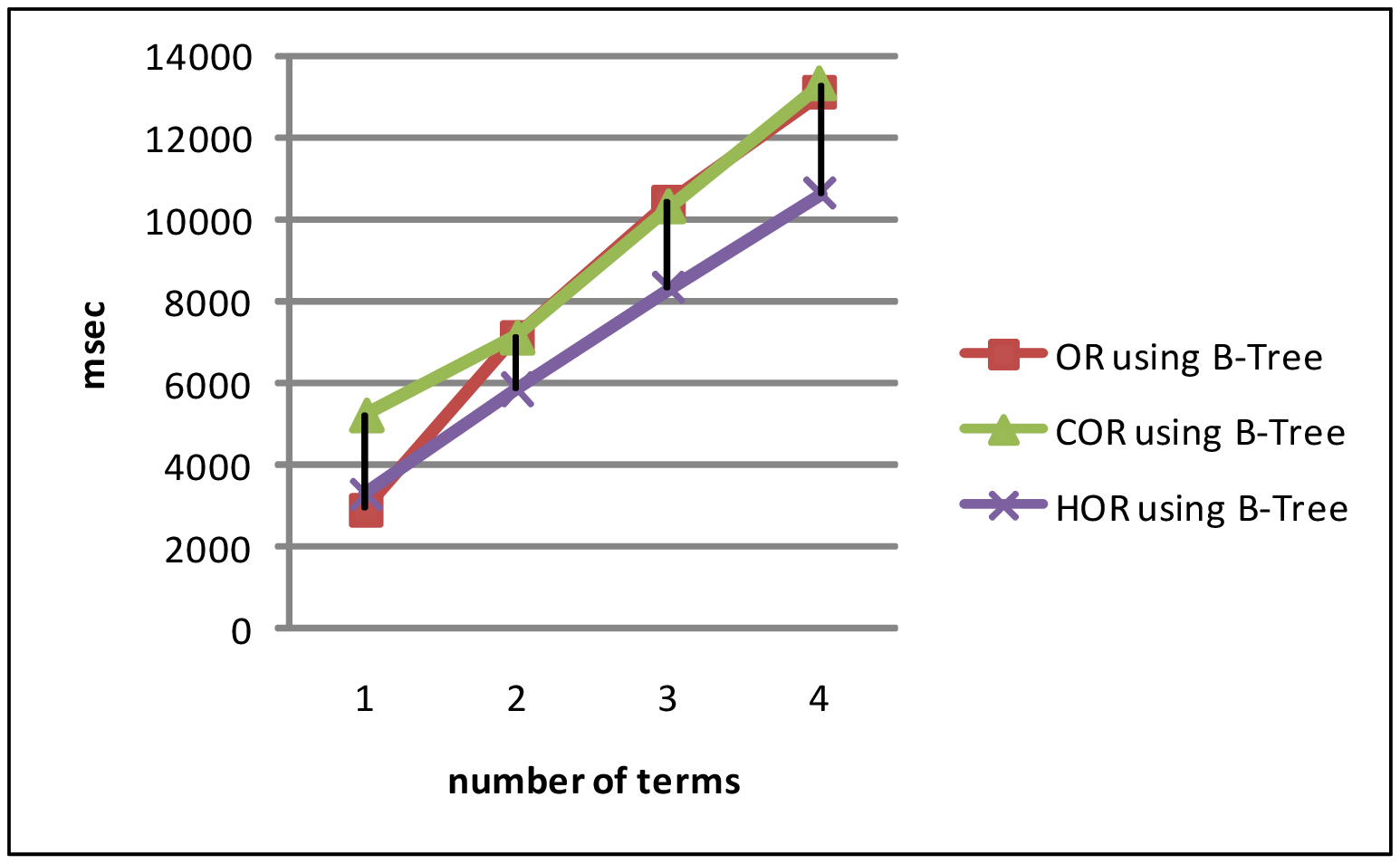}}
     \hspace{.1in}
     \subfigure[Query Evaluation using $Hash$]{
          \label{fig:queryEvaluationHash}
          \includegraphics[width=56mm]{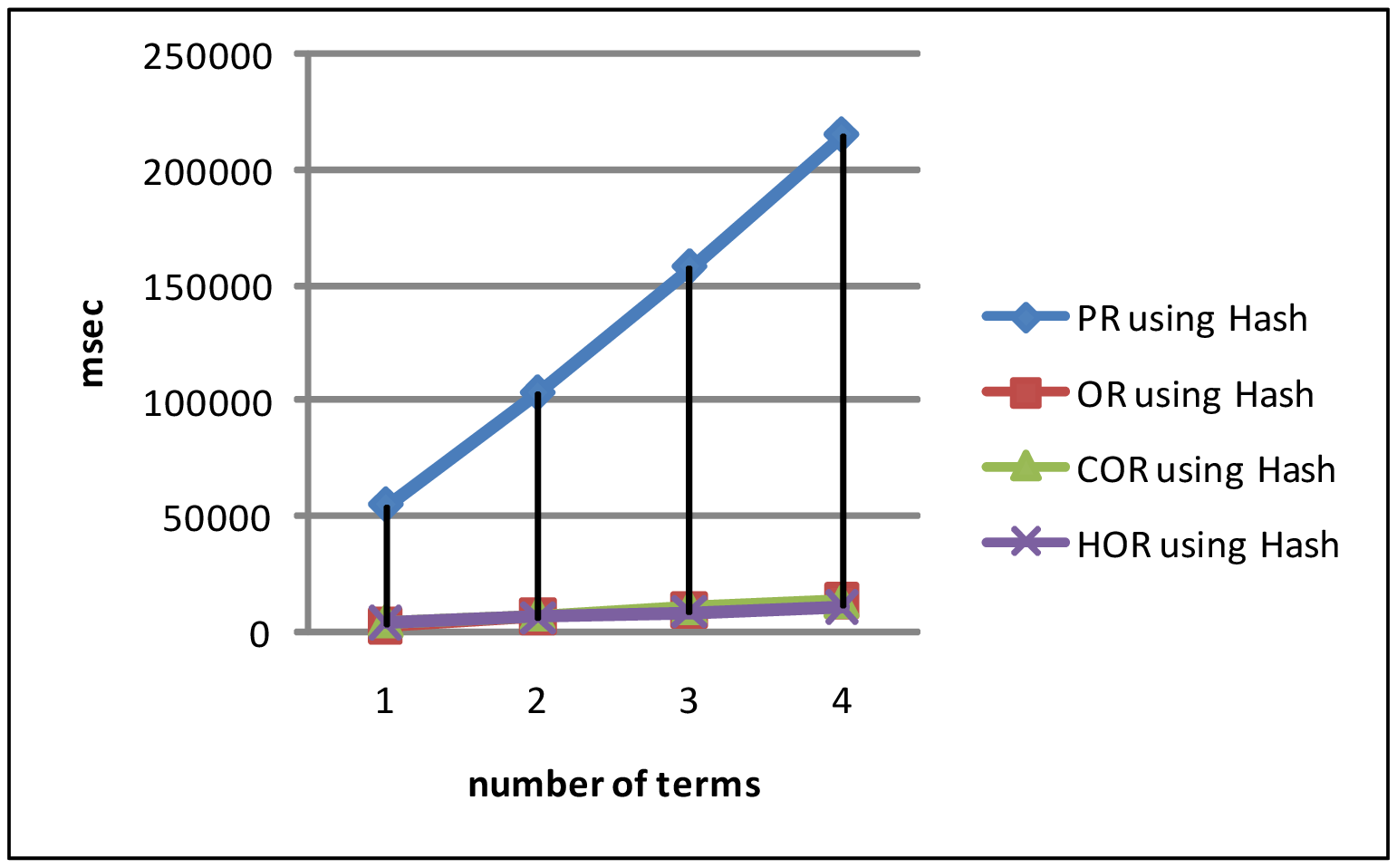}}
    \caption{Index Creation and Query Evaluation Times}
\end{figure}

In this paper we proposed and evaluated four different ORDBMS representations
for text indexing.
\ORIF\ representations were found to be the most efficient,
being one order of magnitude less space consuming
and more than 20 times faster in query evaluation
compared to \ROne.
Specifically,
for a collection of 1 million documents occupying 200 GB,
\ROne\ needs almost 3 days to copy the tables,
while \ORIF\ representations need 30 minutes or less.
The \ROne-index occupies 10.5 GB, while the rest
representations need only 500 MB.
This means that the \ORIF\ index 
is the 0.25\% of the collection size.
Figures \ref{fig:sizes},
\ref{fig:indexCopyTime},
\ref{fig:dbCreationTime},
\ref{fig:queryEvaluationBTree} and
\ref{fig:queryEvaluationHash} summarize the results of the experimental evaluation.
It is worth mentioning
that almost all previous related works
(e.g. \cite{Grossman95aparallel,502655,TopX})
adopt a \ROne-like representation.
As there are numerous applications based on ORDBMS, our
findings can be exploited
for enriching these applications with more scalable IR capabilities.
To avoid misunderstandings,
we do not suggest the adoption of databases instead of inverted indices, we
just identified ways to speedup DBMS-based text indices.

Although some DBMSs currently provide tuple ranking based on text-valued attributes
(e.g. Oracle 9i Text extension, postgreSQL Full Text Search, etc),
an implementation over these services does not allow supporting different retrieval models.
Instead the ranking would be
tightly coupled with the peculiarities of the particular DBMS.
For this reason we based query evaluation
on a small set of elementary queries that enable
implementing  several retrieval models in a flexible manner.
However,
an alternative
approach
would be to use  fewer and more complex  SQL queries that could even compute the
 ranked set of objects in one shot
(depending on the retrieval model).
This is an additional issue for further research.
Furthermore, we plan
to compare the DBMS approach with the classical inverted file approach on the same collection,
and to
compare the efficiency of $B^+Tree$ indices
with the $tree-Trie$ index \cite{treeTrie} that has been proposed to index relationships
with set-value attributes.
Finally, and in order to optimize document based access on the \RFour\ representation,
we plan to evaluate $GIN$ indices on top of the \emph{hstore} values. 
The evaluation of the efficiency in case of concurrent queries,
as well as the investigation of the applicability
of parallelization techniques (e.g. map-reduce) over a DBMS-index,
are subject for future research.

\bibliographystyle{plain}
\bibliography{bib/paper}


\end{document}